\documentclass[prd,twocolumn,eqsecnum,showpacs,amsfonts,amssymb]{revtex4}

\usepackage{graphicx}
\usepackage{rotating}

\usepackage{bm}

\newcommand{\beq}{\begin{equation}}
\newcommand{\eeq}{\end{equation}}
\newcommand{\beqs}{\begin{eqnarray}}
\newcommand{\eeqs}{\end{eqnarray}}

\begin{document}

\title{An Integral Formalism for the Construction of Scheme 
Transformations in Quantum Field Theory} 

\author{Gongjun Choi and Robert Shrock}

\affiliation{C. N. Yang Institute for Theoretical Physics \\
Stony Brook University, Stony Brook, NY 11794 }

\begin{abstract}

We present an integral formalism for constructing scheme transformations in
a quantum field theory. We apply this to generate several new useful scheme 
transformations. A comparative analysis is given of these scheme 
transformations in terms of their series expansion coefficients and their
resultant effect on the interaction coupling, in particular at a zero of the
beta function away from the origin in coupling-constant space. 

\end{abstract}


\maketitle


\section{Introduction}
\label{intro}

The dependence of the interaction coupling in a quantum field theory on the
Euclidean momentum scale, $\mu$, where it is probed, is of basic importance.
This is determined by the beta function of the theory \cite{rg}.  For
simplicity, we focus here on a theory (in four spacetime dimensions at zero
temperature) with only one dimensionless interaction coupling . There has long
been interest in a possible zero of the beta function away from the origin in
coupling-constant space. For an infrared-free theory such as quantum
electrodynamics or $\lambda \phi^4$ this would be an ultraviolet fixed point
(UVFP) of the renormalization group (RG), while for an asymptotically free
non-Abelian gauge theory, this would be an infrared fixed point (IRFP) of the
renormalization group, calculated to a given order in perturbation theory, in
both cases.  Let us consider the latter case, of a non-Abelian gauge theory
with a simple gauge group and hence a single gauge coupling.  We shall denote
the running gauge coupling as $g \equiv g(\mu)$ and define
$\alpha(\mu)=g(\mu)^2/(4\pi)$.  For technical simplicity, we take the fermions
to be massless and avoid inclusion of any scalar fields, so that the theory
involves only one dimensionless interaction coupling. With a given fermion
content, the theory possesses an IRFP at the two-loop level if the two-loop
coefficient in the beta function, $b_2$, has a sign opposite to that of the
one-loop coefficient, $b_1$ (see Eq. (\ref{beta}) below).  At the two-loop
($2\ell$) level, this IRFP occurs at the value $\alpha=\alpha_{IR,2\ell}=-4\pi
b_1/b_2$.  It is clearly desirable to calculate the value of this IR zero of
the beta function to higher-loop order to achieve greater accuracy in its
determination.  However, while the one-loop and two-loop coefficients in the
beta function are independent of the scheme used for regularization and
renormalization, the coefficients at the level of three loops and higher depend
on this scheme \cite{gross75}. Indeed, this scheme dependence of higher-loop
calculations is a general property of quantum field theories.

It is therefore incumbent upon one to assess how sensitive a given quantity is
to the scheme used for the higher-loop calculation of this quantity.  Here we
concentrate on the calculation of the location of a zero of a beta function
away from zero coupling but still at sufficiently small coupling that one can
use perturbative methods.  A procedure to assess the scheme dependence of the
location of this zero in the beta function is to carry out the calculation
first in a given scheme, obtain a result for the value of the IR zero at
$n$-loop ($n\ell$) order, $\alpha_{IR,n\ell}$, then apply a scheme
transformation, calculate the zero to this order in the transformed scheme,
denoted $\alpha_{IR,n\ell}'$, and determine the fractional shift in the value.
In a series of papers this program has been implemented \cite{scc}-\cite{schl}.
Refs. \cite{scc,sch} pointed out that it is significantly more difficult to
construct scheme transformations that can be applied away from the origin in
coupling-constant space than it is to construct such transformations that are
applicable in the vicinity of the origin, such as those used in quantum
chromodynamics (QCD) calculations in the perturbative region, i.e., for 
small $\alpha_s$.  For example, consider the scheme transformation
\beq
\alpha = \frac{1}{2}\tanh(2\alpha') \ . 
\label{st_tanh_r8pi}
\eeq
This is perfectly well-behaved near zero coupling,  
$\alpha=\alpha'=0$, where it approaches the identity transformation, but is
unacceptable at a generic zero of the beta function. This is clear from the
inverse transformation, which is
\beq
\alpha'=\frac{1}{4} \ln \Big ( \frac{1+2\alpha}{1-2\alpha} \Big ) \ . 
\label{st_tanh_inverse_r8pi}
\eeq
As $\alpha$ approaches the value 1/2 from below, $\alpha' \to \infty$, and for
$\alpha > 1/2$, $\alpha'$ is complex.  

Since, as was noted, coefficients in the beta function at the level of three
loops and higher are scheme-dependent, it was anticipated that, in the vicinity
of zero coupling, as in QCD, one could transform to a scheme where these
coefficients vanish \cite{thooft77}. Another important result from
\cite{scc}-\cite{sch3} was an explicit construction of a scheme transformation
that removes terms of loop level $n \ge 3$ in the beta function in the vicinity
of $\alpha=\alpha'=0$ and the demonstration that it is much more difficult to
try to carry out this removal of higher-loop terms at a larger but still
perturbative value of the coupling away from the origin.  In \cite{sch3} a
generalized scheme transformation denoted $S_{R,m,k_1}$ with $m \ge 2$ was
presented with the property that it eliminates the $n$-loop terms in the beta
function of a gauge theory from loop order $n=3$ to order $n=m+1$, inclusive
and can be optimized to perform this removal in a substantial range of
couplings away from the origin.

There is thus a need to construct and apply scheme transformations that are
applicable not only near the origin in coupling-constant space (where they
automatically reduce to the identity), but also at a zero of the beta function
located away from the origin.  The previous works \cite{scc}-\cite{schl}
addressed this task and studied applications at an IR zero of the beta function
in an asymptotically free non-Abelian gauge theory.  Early interest in such a
zero had made use of the scheme-independent one-loop and two-loop coefficients
and had noted the associated behavior of scaling with anomalous dimensions
\cite{b1,b2}. Later, it was observed that in the region where the number of
fermions approaches the maximum value allowed by asymptotic freedom (the value
where the one-loop coefficient $b_1$ vanishes), this IR zero occurs at small
coupling \cite{bz}.  Moving away from this region toward larger values of
$\alpha_{IR,2\ell}$ requires higher-order calculations \cite{gk}-\cite{bc} to
achieve reasonable accuracy, whence the necessity of dealing with the issue of
scheme dependence.  These calculations made use of expressions for the
three-loop and four-loop beta function coefficients, $b_3$ \cite{b3} and $b_4$
\cite{b4} that had been calculated in the $\overline{\rm MS}$ scheme
\cite{msbar}. Refs. \cite{bvh}-\cite{bc} carried out this analysis for a
general gauge group and for fermions in both the fundamental representation and
in the adjoint and rank-2 tensor representations. A particularly powerful
approach uses scheme transformations that are dependent on an auxiliary
parameter, $r$, with the property that as $r \to 0$, they approach the
identity; by varying $r$ continuously away from $r=0$, one can thus study the
scheme dependence as a function of this continuous variable
\cite{scc}-\cite{schl}.  A valuable result from this program of higher-order
perturbative computations of the values of IR fixed points of asymptotically
free non-Abelian gauge theories is improvement in the accuracy of calculations
of anomalous dimensions, such as the anomalous dimension of the fermion
bilinear operator, evaluated at the IR fixed point, $\gamma_{IR,n\ell} \equiv
\gamma_{n\ell}(\alpha_{IR,n\ell})$. The result can then be compared with
lattice calculations that are fully nonperturbative in the gauge coupling,
although involving other approximations, such as finite lattice spacing, finite
lattice volume, removal of fermion doubler modes, etc. \cite{lgt}.  For example
for a (vectorial) SU(3) gauge theory with $N_f=12$ massless Dirac fermions, the
values of $\gamma_{IR,n\ell}$ at the two-loop, three-loop, and four-loop level
were found to be 0.773, 0.312, and 0.253, respectively.  The four-loop value is
in good agreement with the lattice calculations $\gamma_{IR} = 0.27 \pm
0.03$ \cite{hasenfratz1}, $\gamma_{IR} \simeq 0.25$ \cite{hasenfratz2}, and
$\gamma_{IR}=0.235 \pm 0.046$ \cite{lombardo}.  This shows the value of
calculating $\alpha_{IR,n\ell}$ to higher-loop order, since one evaluates 
$\gamma_{n\ell}$ at $\alpha=\alpha_{IR,n\ell}$ to obtain $\gamma_{IR,n\ell}$. 
Another approach to ascertaining the degree of scheme dependence is to use 
different schemes, such as the modified minimal subtraction 
$\overline{\rm MS}$ \cite{msbar}, momentum subtraction (MOM) \cite{mom}, and
${\rm RI}'$ \cite{rip}, for the calculation of $\alpha_{IR,n\ell}$ 
and then compare the results \cite{tr,gracey2015} (see also \cite{tr2016}).  

The program in \cite{scc}-\cite{schl} is complementary to work on optimized
schemes to be applied in the neighborhood of the origin, as in perturbative QCD
calculations \cite{mom,early,brodsky}.  Scheme transformations have also been
used in recent studies of possible UV zeros in a beta function for several
types of non-asymptotically free theories, including a U(1) gauge theory
\cite{lnf} and a globally invariant O($N$) $\lambda |{\vec \phi}|^4$ theory
\cite{lam}.  We do not explicitly consider supersymmetric field theories here
but note that scheme transformations have also been studied in such theories
(e.g., \cite{susyschemes}-\cite{nsvz}).

In this paper we report important further progress in this program of
constructing scheme transformations that are acceptable for applications away
from, as well as near, the origin in coupling-constant space.  We present a
method for generating scheme transformations based on an integral
formalism.  We demonstrate the usefulness of this integral formalism by
utilizing it to construct several new scheme transformations that can be
applied to determine the degree of scheme dependence of a higher-loop
calculation of a zero of the beta function away from the origin in
coupling-constant space. We also present a comparative analysis of scheme 
transformations in terms of the coefficients that enter in their Taylor series
expansions in the relevant coupling and use this to infer how they shift a 
coupling that is in the perturbative region. 

This paper is organized as follows.  In Sect. \ref{beta_section} we discuss
some relevant background on the beta function.  In
Sect. \ref{scheme_transformation_section} we give a general discussion of
scheme transformations, including the set of acceptable conditions that they
must satisfy. In this section we derive a basic property concerning how a
scheme transformation shifts the value of the coupling. In
Sect. \ref{integral_formalism_section} we present our new integral formalism
for the construction of acceptable scheme transformation. In the subsequent
sections we apply this formalism to generate a number of new useful scheme
transformations for which explicit inverses can be calculated.  Some
comparative comments are included in Sect. \ref{comparison_section}, and our
conclusions are given in Sect. \ref{conclusions}.


\section{Beta Function} 
\label{beta_section} 

Here we briefly mention some necessary background for our later discussion. 
As noted before, although our results are more general, we shall focus in this
paper on a non-Abelian gauge theory with a simple gauge group $G$ and running 
gauge coupling $g(\mu)$, with a fermion content chosen such that the theory is
asymptotically free.  Such theories have
the appeal that there is at least one regime, namely large Euclidean
energy/momentum $\mu$ in the deep UV, where one can carry out reliable
perturbative calculations.  We define 
\beq
a(\mu) \equiv \frac{g(\mu)^2}{16\pi^2} = \frac{\alpha(\mu)}{4\pi} \ .
\label{a}
\eeq
The argument $\mu$ will often be suppressed in the notation. 
The beta function is $\beta_g = dg/dt$ or
equivalently,
\beq
\beta_\alpha \equiv \frac{d\alpha}{dt} = \frac{g}{2\pi} \, \beta_g \ ,
\label{betadef}
\eeq
where $dt=d\ln \mu$. This function has the series expansion
\beq
\beta_\alpha = -2\alpha \sum_{\ell=1}^\infty b_\ell \, a^\ell \ ,
\label{beta}
\eeq
where an overall minus sign has been extracted in the prefactor. 
The $n$-loop ($n\ell$) beta
function, denoted $\beta_{\alpha,n\ell}$, is given by (\ref{beta}) with 
the upper limit on the $\ell$ summation taken to be $n$ rather than
$\infty$.  If the theory has an IR zero of the $n$-loop beta function 
$\beta_{\alpha,n\ell}$, we denote it by 
$\alpha_{IR,n\ell} = 4\pi a_{IR,n\ell}$. 

As the reference scale $\mu$ decreases from large values in the deep UV to
smaller scales toward the IR and $\alpha(\mu)$ increases, it approaches the
value at the IR zero of the beta function, which we denote generically as
$\alpha_{IR}$ in this paragraph.  If the gauge group and fermion content are
such that $\alpha_{IR}$ is sufficiently small, then the theory evolves to a
chirally symmetric phase in the IR and $\alpha \to \alpha_{IR}$ as $\mu \to 0$,
so that $\alpha_{IR}$ is an exact IRFP of the renormalization group.  If, on
the other hand, $\alpha_{IR}$ is sufficiently large, then the gauge interaction
produces bilinear fermion condensate(s) and associated spontaneous chiral
symmetry breaking.  In this case, the fermions pick up dynamical masses and are
integrated out of the low-energy effective field theory that is operative at
scales below the scale of the condensate formation.  Hence, in this low-energy
theory, the beta function changes form, and $\alpha(\mu)$ evolves away from
$\alpha_{IR}$ toward stronger coupling.  In this case, $\alpha_{IR}$ is only an
approximate IRFP of the renormalization group. 


\section{Scheme Transformations} 
\label{scheme_transformation_section}

A scheme transformation is a mapping between $\alpha$ and
$\alpha'$ or equivalently, between $a$ and $a'$, namely 
\beq
a = a'f(a') \equiv F(a') \ . 
\label{aap}
\eeq
Here it is convenient to introduce the notation $F(a')$ to emphasize the
functional dependence of $a$ on $a'$.  
In the limit where $a$ and $a'$ vanish, the theory becomes free, so a scheme
transformation has no effect, i.e., it should approach the identity.  
This implies that 
\beq
f(0) = 1 \ . 
\label{c0}
\eeq

The functions $f(a')$ that we consider have Taylor series expansions about 
$a=a'=0$ of the form
\beq
f(a') = 1 + \sum_{s=1}^{s_{max}} k_s (a')^s \ , 
\label{fapseries}
\eeq
where the coefficients $k_s$ are constants.  Here, $s_{max}$ may be
finite or infinite.  Thus, these functions $f(a')$ automatically satisfy the 
condition (\ref{c0}). Equivalently, 
\beq
a = F(a') = a' + \sum_{s=1}^{s_{max}} k_s (a')^{s+1} \ . 
\label{Fapseries}
\eeq

By using the method of reversion of series \cite{ww}, one can calculate 
a Taylor series expansion for the inverse scheme transformation, 
$a'=F^{-1}(a)$, from the series (\ref{Fapseries}).  This series for the inverse
may be written as 
\beq
a' = F^{-1}(a) = a + \sum_{s=1}^{s_{max}} \rho_s \, a^{s+1} \ . 
\label{Fap_inverse_series}
\eeq
In terms of the $k_s$ coefficients, we have 
\beq
\rho_1 = -k_1 \ , 
\label{rho1}
\eeq
\beq
\rho_2 = 2k_1^2-k_2 \ , 
\label{rho2}
\eeq
\beq
\rho_3 = 5k_1k_2 - k_3 - 5k_1^3 \ , 
\label{rho3}
\eeq
\beq
\rho_4 = 6k_1k_3+3k_2^2+14k_1^4-k_4-21k_1^2k_2 \ , 
\label{rho4}
\eeq
and so forth for higher $s$. 

Since $a$ and $a'$ are small in the perturbative region where these scheme
transformations are applicable, it is of interest to consider truncations of
the series (\ref{Fapseries}) and (\ref{Fap_inverse_series}).  At the lowest
order beyond the identity, Eq. (\ref{Fapseries}) reduces
to the equation $a=a'(1+k_1 a')$. Although this is a quadratic equation for
$a'$, which has two formal solutions, only one is physical, as uniquely
determined by the requirement that it must reduce to the identity as $k_1 \to
0$.  This solution is
\beq
a' = \frac{1}{2k_1} \Big [ -1 + \sqrt{1+4k_1a} \ \Big ] \ . 
\label{st_lowest_order_sol}
\eeq
Similarly, to the same order, the series for the inverse transformation,
Eq. (\ref{Fap_inverse_series}), reduces to $a'=a(1-k_1a)$, and in the same way,
although this is a quadratic equation in $a$ with two formal solutions, one
uniquely determines the physical solution by the requirement that as 
$k_1 \to 0$, it reduces to the identity.  This solution is
\beq
a = \frac{1}{2k_1}\Big [ 1 - \sqrt{1-4k_1a'} \ \Big ] \ . 
\label{st_inverse_lowest_order_sol}
\eeq
As is evident from either Eq. (\ref{st_lowest_order_sol}) or 
(\ref{st_inverse_lowest_order_sol}), for small $a$, 
if $k_1 > 0$, then $a > a'$, while if $k_1 < 0$, then $a < a'$. 

We next give a general inequality that determines whether a scheme
transformation increases or decreases the value of the coupling for small $a$,
as a function of the sign of the lowest nonzero coefficient $k_s$ in
Eq. (\ref{Fapseries}).  This inequality applies even if this lowest nonzero
coefficient is not $k_1$.  It is useful, since some scheme transformations and
their inverses have respective Taylor series (\ref{Fapseries}) and
(\ref{Fap_inverse_series}) in which $k_1=0$.  This is the case, for example,
with the transformations (\ref{st_sinh}) and (\ref{st_tanh}) below. Let us
denote the lowest-order nonzero coefficient $k_s$ in Eqs. (\ref{fapseries}) and
(\ref{Fapseries}) as $k_{s_{min}}$.  Then we find the following general
inequality for small $a$ (and hence also small $a'$),
\beqs
& & k_{s_{min}} > 0 \ \Longrightarrow \ a > a' \ , 
\cr\cr
& & k_{s_{min}} < 0 \ \Longrightarrow \ a < a' \ . 
\label{ks_aaprel}
\eeqs

A number of the scheme transformations studied in \cite{scc}-\cite{schl} depend
on a parameter (denoted $r$ in these works) and hence are actually
one-parameter families of scheme transformations. Here and below, we shall
often refer to a one-parameter family of scheme transformations as a single
scheme transformation, with the dependence on the parameter $r$ taken to be
implicit. In accordance with the series expansion (\ref{Fapseries}), 
$F(a')$ has the property
\beq
F'(0) \equiv \frac{dF(a')}{da'}\Big |_{a'=0} = 1 \ . 
\label{dFdap1}
\eeq
(No confusion should result from the prime used here for differentiation and
the prime on $a'$, which does not indicate any differentiation but just
distinguishes $a'$ from $a$.)

From (\ref{fapseries}), it follows that the Jacobian
\beq
J = \frac{da}{da'}= \frac{d\alpha}{d\alpha'} 
\label{j}
\eeq
can be expanded as 
\beq
J = 1 + \sum_{s=1}^{s_{max}} (s+1)k_s  \, (a')^s
\label{jseries}
\eeq
and therefore satisfies the condition 
\beq
J=1 \quad {\rm at } \ \ a=a'=0 \ . 
\label{jacobianazero}
\eeq
Since $J$ is the derivative $da/da'$, it is naturally expressed as a function
of either $a'$ or $a$.

The beta function in the transformed scheme is 
\beq
\beta_{\alpha'} \equiv \frac{d\alpha'}{dt} = \frac{d\alpha'}{d\alpha} \, 
\frac{d\alpha}{dt} = J^{-1} \, \beta_{\alpha} \ , 
\label{betaap}
\eeq
with the series expansion 
\beq
\beta_{\alpha'} = -2\alpha' \sum_{\ell=1}^\infty b_\ell' \, (a')^\ell  \ . 
\label{betaprime}
\eeq
Owing to the fact that Eqs. (\ref{betaap}) and (\ref{betaprime}) refer to the
same function, one can solve for the $b_\ell'$ in terms of the $b_\ell$ and
$k_s$.  This yields the known results $b_1'=b_1$ and $b_2'=b_2$ for the
one-loop and two-loop coefficients.  In Refs. \cite{scc,sch}, explicit
expressions were calculated and presented for higher-loop coefficients
$b_\ell'$ with $\ell \ge 3$ in terms of the $b_\ell$ and $k_s$.

To be physically acceptable, a scheme transformation must satisfy
several conditions, as was discussed in \cite{sch}. We state these for an
asymptotically free gauge theory: (i) condition $C_1$: the scheme
transformation must map a real positive $\alpha$ to a real positive $\alpha'$;
(ii) $C_2$: the scheme transformation should not map a moderate value of
$\alpha$, for which perturbation theory may be reliable, to a value of
$\alpha'$ that is so large that perturbation theory is unreliable, or vice
versa; (iii) $C_3$: the Jacobian $J$ should not vanish (or diverge) or else the
transformation would be singular; and (iv) $C_4$: since the existence of an IR
zero of $\beta$ is a scheme-independent property of an theory, a scheme
transformation must satisfy the condition that $\beta_\alpha$ has an IR zero if
and only if $\beta_{\alpha'}$ has an IR zero. Since $J=1$ for $a=a'=0$, 
the condition $C_3$ implies that $J$ must be positive.  Clearly, these 
apply both to a scheme transformation from $a$ to $a'$ and to the inverse from
$a'$ to $a$. 

These four conditions $C_1$-$C_4$ can always be satisfied by scheme
transformations used to study the UV fixed point in an asymptotically free
theory.  This is clear from the fact that $f(a')$ approaches 1 as $a' \to 0$ in
(\ref{fapseries}), so the transformation approaches the identity in this limit.
However, as was pointed out in \cite{scc} and shown with a number of
examples in \cite{scc}-\cite{sch3}, they are not automatically satisfied, and
indeed, are quite restrictive conditions when one applies the scheme
transformation at a zero of the beta function away from the origin, $\alpha=0$,
i.e., at an IR zero of the beta function for an asymptotically free theory or a
possible UV zero of the beta function for an infrared-free theory.  

Some further remarks on the applicability of a scheme transformation are
appropriate here.  Since a major application of scheme transformations is to
determine how sensitive the value of a zero of the beta function, calculated to
loop order $n=3$ or higher, is to the scheme used for the calculation, and
since such a calculation is only reliable if the coupling $\alpha$ is not too
large, it follows that one need only impose the conditions C$_1$-C$_4$ in 
this range of values of $\alpha$ that are not so large as to render
perturbative calculations inapplicable.  Nevertheless, it is valuable to have a
scheme transformation that satisfies all of the conditions C$_1$-C$_4$ for
arbitrary (physical, i.e., real, positive) values of $\alpha$, so that one does
not have to be concerned about trying to choose some nominal value of $\alpha$
beyond which it cannot be applied.  To expand upon this point, we may compare
and contrast two illustrative scheme transformations \cite{sch}.  One
of these satisfies the conditions C$_1$-C$_4$ for arbitrary values of $\alpha$.
This is the transformation
\beq
a = F(a') = \frac{1}{r} \sinh(ra')
\label{st_sinh}
\eeq
with inverse
\beq
a' = \frac{1}{r}\ln\Big [ ra + \sqrt{1+(ra)^2} \ \Big ] 
\label{st_sinh_inverse}
\eeq
and Jacobian, expressed equivalently as a function of $a'$ and $a$,
\beq
J = \cosh(ra') = \sqrt{1+(ra)^2} \ . 
\label{j_sinh}
\eeq
This is an example of a class of one-parameter families of scheme
transformations whose members are invariant under reversal in sign of the
auxiliary parameter $r$.  Hence, for such transformations, we can, without loss
of generality, take this parameter $r$ to be nonnegative, and, as in
\cite{sch}, we shall do so.  The application of this transformation in
\cite{sch} to the IR zero in the beta function in an SU($N$) theory with $N_f$
fermions in the fundamental representation showed that for moderate $r$ and for
values of $\alpha_{IR,n\ell}$ for $n=3$ and $n=4$ loops that were not too
large, these values were not sensitively dependent on the scheme used for their
calculation.

In contrast, consider the scheme transformation
\beq
a = F(a') = \frac{1}{r} \tanh(ra')
\label{st_tanh}
\eeq
with the inverse
\beq
a' = \frac{1}{2r} \ln \Big ( \frac{1+ra}{1-ra} \Big ) 
\label{st_tanh_inverse}
\eeq
and Jacobian, expressed as a function of $a'$ and, equivalently, of $a$: 
\beq
J = \frac{1}{\cosh^2(ra')} = (1-ra)(1+ra) \ . 
\label{j_tanh}
\eeq
Again, we may, without loss of generality, take $r$ to be nonnegative.
Evidently, the inverse transformation (\ref{st_tanh_inverse}) and the Jacobian
are singular at $a=1/r$, i.e., $\alpha=4\pi/r$. The transformation
(\ref{st_tanh_inverse}) thus does not satisfy the conditions C$_1$-C$_4$ for
arbitrary values of $a$. Indeed, a special case of this transformation with
$r=8\pi$ was given above in Eq. (\ref{st_tanh_r8pi}) and the singularity at
$\alpha=1/2$ in the inverse, Eq. (\ref{st_tanh_inverse_r8pi}) was noted. Hence,
the scheme transformation (\ref{st_tanh}) is not as well-behaved as
(\ref{st_sinh}) is.  However, if one restricts the parameter $r$ to
sufficiently small values that the singularity at $\alpha=4\pi/r$ occurs at a
value of $\alpha$ substantially greater than unity, where one would not try to
use perturbative methods, then this singularity would not prevent one from
utilizing this transformation.


\section{Integral Formalism for Construction of Scheme Transformations}
\label{integral_formalism_section}

Here we introduce and apply a general integral formalism for the construction
of one-parameter families of scheme transformations.  In this formalism, the 
starting point is a choice of a Jacobian $J(y)$ that will be used 
as the integrand of an integral representation of the function $F(a')$ defined
in Eq. (\ref{aap}): 
\beq
a = F(a') = \int_0^{a'} J(y)\, dy \ . 
\label{fjint}
\eeq
We choose $J(y)$ to be an analytic function of $y$ satisfying the 
condition
\beq
J(0)=1 \ . 
\label{j01}
\eeq
This guarantees that $J(a')$ and $f(a')$ have the respective Taylor series
expansions (\ref{jseries}) and (\ref{fapseries}) and hence that $f(a')$
satisfies the condition (\ref{c0}).  As discussed above, the condition C$_3$
for an acceptable scheme transformation is that the Jacobian must not vanish,
since otherwise the transformation is singular.  The property $J(0)=1$ together
with analyticity of $J$ imply that $J$ must be positive for the ranges of
couplings $a$ and $a'$ that are relevant for perturbative calculations for
which these scheme transformations are applicable. Thus, we require that $J(y)
> 0$ throughout the range of the integration variable $y$ in Eq. (\ref{fjint}).
We can also include dependence of the scheme transformation on a (real)
auxiliary parameter, denoted $r$. Differentiating Eq. (\ref{fjint}) and using a
basic theorem from calculus (Eq. (\ref{diffint}) in Appendix
\ref{diffint_appendix}) yields the relation $dF(a')/da' = da/da'= J(a')$, in
agreement with Eq. (\ref{j}).  Using an appropriate choice for the Jacobian
$J(z)$, we can also satisfy conditions C$_1$-C$_4$.

In addition to these general conditions for the acceptability of a scheme
transformation, another important aspect of the analysis is the ease of
inverting the transformation to solve for $a'$ from $a$.  As was evident in
Refs. \cite{sch}-\cite{sch3}, for algebraic scheme transformations with finite
values of $s_{max}$ in Eq. (\ref{fapseries}), the inversion required the
solution of an algebraic equation and a choice of which root to take for this
solution.  In contrast, for cases of algebraic or transcendental scheme
transformations with $s_{max}=\infty$, the inverse transformations were often
simpler, in the sense that one did not have to make such a choice of which root
of an algebraic equation to take.

To show the usefulness of this integral formalism for the construction of
acceptable scheme transformations, we will employ it to generate a number of
new scheme transformations which also have the advantage that their inverses
can be calculated explicitly in closed form.  Before doing this, we first
illustrate how the method works with some scheme transformations that have
already been studied in \cite{scc}-\cite{schl}, which we showed to be
acceptable for the analysis of a zero in a beta function located away from the
origin in coupling constant space, in particular, an IR zero of the beta
function of an asymptotically free non-Abelian gauge theory. Let us consider,
for example, the scheme transformation (\ref{st_sinh}) studied in \cite{sch}.
To show how one could use our present integral formalism to construct this
scheme transformation, we start with $J$ and replace the variable $a'$ by the
integration variable $y$ to get $J(y)=\cosh(ry)$.  Substituting this function
into Eq. (\ref{fjint}), we obtain
\beq 
a = F(a') = \int_0^{a'} \cosh(ry) \, dy = \frac{1}{r} \sinh(ra') \ , 
\label{fasinh}
\eeq
thereby rederiving the transformation (\ref{st_sinh}). 

Other examples are provided by the scheme transformations that we
studied in \cite{schl}.  One of these is 
\beq
a = F(a') = \frac{1}{r} \ln(1+ra')
\label{slr}
\eeq
with inverse
\beq
a' = \frac{e^{ra}-1}{r} \ . 
\label{aap_slr}
\eeq
The Jacobian, expressed equivalently as a function of $a'$ and $a$, is 
\beq
J = \frac{1}{1+ra'} = e^{-ra} \ . 
\label{j_slr}
\eeq
Again replacing the variable $a'$ by $y$ to get $J(y)=1/(1+ry)$ and then
substituting this into Eq. (\ref{fjint}), we reproduce
the original transformation: 
\beq
a = F(a') = \int_0^{a'} \frac{dy}{1+ry} = \frac{1}{r}\ln(1+ra') \ . 
\label{Fapslr}
\eeq
Here, as we discussed in \cite{schl}, the parameter $r$ is restricted to lie in
the range $r > -1/a'$ to avoid a singularity in the transformation and is
further restricted by the condition that the scheme transformation satisfies
conditions C$_1$-C$_4$.  Similarly, if one uses $J(y)=1/(1-ry)^2$ in
Eq. (\ref{fjint}), one obtains another scheme transformation considered in
\cite{schl}, namely
\beq
a = F(a') = \frac{a'}{1-ra'} \ . 
\label{st_rational}
\eeq
%


\section{Transformation with an Algebraic $J(y)$}
\label{algebraic_power_section}

We next proceed to present new scheme transformations that we have constructed
using our integral formalism.  Recall that the starting point for the procedure
is a choice of the Jacobian function $J(y)$ that serves as the integrand in
Eq. (\ref{fjint}) and that satisfies the requisite conditions that it is
analytic and that $J(0)=1$.  For our first new transformation, we choose a
$J(y)$ of algebraic form, namely 
\beq
J(y) = (1+ry)^p \ , 
\label{j1alg}
\eeq
where the power $p$ is a positive real number. Then, calculating the 
integral in Eq. (\ref{fjint}), we obtain the scheme transformation 
\beq
a = F(a') = \frac{(1+ra')^{p+1}-1}{r(p+1)} \ . 
\label{F_alg}
\eeq
The resultant series expansion for $f(a')=F(a')/a'$ has the form of
Eq. (\ref{fapseries}) with 
\beq
k_s = \frac{r^s}{(s+1)} \, {p \choose s} = 
\frac{r^s}{(s+1)!} \, \prod_{\ell=0}^{s-1}(p-\ell) \ ,
\label{ks_alg}
\eeq
where ${a \choose b} = \ a!/[b!(a-b)!]$ is the binomial coefficient.  This is a
finite series if $p$ is an integer, and an infinite series otherwise.  We list
the coefficients $k_s$ explicitly for the first few values of $s$ for this
scheme transformation and for others discussed in this paper in Table
\ref{ks_table}. The series (\ref{ks_alg}) has $s_{min}=1$ and
\beq
k_{s_{min}} = k_1 = \frac{p \, r}{2} \ . 
\label{k1_alg}
\eeq

The inverse transformation is 
\beq
a' = \frac{1}{r} 
\Big [ \Big \{ (p+1)ra+1 \Big \}^{\frac{1}{p+1}} - 1 \Big ] \ . 
\label{ap_alg}
\eeq
Using this inverse transformation, one can express the Jacobian equivalently as
a function of $a'$: 
\beq
J=(1+ra')^p = \Big [ (p+1)ra + 1 \Big ]^{\frac{p}{p+1}} \ . 
\label{j_alg}
\eeq
There are two immediate restrictions on the parameter $r$ arising from the
requirement that $J > 0$ and that there not be any singularity in the scheme
transformation (\ref{F_alg}), namely
\beq
r > -\frac{1}{a'} \quad {\rm and} \quad r > -\frac{1}{(p+1)a} \ . 
\label{rrestrictions_alg}
\eeq
These restrictions are easily met, for example, by requiring that $r$ be
nonnegative. Moreover, the interval in the couplings $\alpha$ where one could
use perturbative calculations reliably only extends up to values 
$\alpha \sim O(1)$, and since $a=\alpha/(4\pi)$, this interval only
extends up to $a \sim O(0.1)$, so for moderate $p$, the lower bounds 
(\ref{rrestrictions_alg}) evaluate to $r \gtrsim -O(10)$.  This lower bound can
easily be satisfied even with moderate negative values of $r$.  With the
restrictions (\ref{rrestrictions_alg}) satisfied, the scheme transformation
(\ref{F_alg}) satisfies the conditions C$_1$-C$_4$.  If $r > 0$, then 
since $k_{s_{min}} > 0$ (where $s_{min}=1$ here) it follows from our general
result (\ref{ks_aaprel}) above that $a > a'$ for small $a, \ a'$. If $r$ is
negative (in the range allowed by above restrictions) then 
$k_1 < 0$, so $a < a'$ for small $a, \ a'$. 


\section{Transformation with a Transcendental $J(y)$}
\label{transcendental_section}

For an application of our integral formalism using a Jacobian that is a 
transcendental function, we choose 
\beq
J(y) = 1+\tanh(ry) \ . 
\label{jtran}
\eeq
Then, doing the integral in Eq. (\ref{fjint}), we obtain 
\beq
a = F(a') = a' + \frac{1}{r} \ln \Big [ \cosh(ra') \Big ] \ . 
\label{F_tran}
\eeq
The resultant series expansion for $f(a')=F(a')/a'$ has the form of
Eq. (\ref{fapseries}) with $k_s=0$ for $s$ even and 
\beq
k_1 = \frac{r}{2}, \quad k_3=-\frac{r^3}{12}, \quad k_5 =\frac{r^5}{45}, \quad
k_7 = -\frac{17r^7}{2520} \ , 
\label{ks_tran}
\eeq
etc. for higher values of $s$.  For comparative purposes, we list these $k_s$
for $s$ up to 4 in Table \ref{ks_table}. 

The Jacobian of this transformation, expressed
as a function of $a'$, is given by Eq. (\ref{jtran}) with $y=a'$. 
The inverse of the scheme transformation has a simple form for 
certain values of $r$.  For example, for $r=1$, the inverse is 
\beq
a' = \frac{1}{2}\ln(2e^a-1) \ . 
\label{ap_tran}
\eeq
Using Eq. (\ref{ap_tran}), one can also express $J$ as a
function of $a$ for this $r=1$ case, obtaining $J=2-e^{-a}$. 

The allowed range of the parameter $r$ is determined by the requirement that
the scheme transformation must satisfy the conditions C$_1$-C$_4$. If $r > 0$,
then, since $k_{s_{min}} > 0$ (where $s_{min}=1$ here), it follows that
$a > a'$ for small $a, \ a'$, while if $r < 0$, then $k_1 < 0$, so
$a < a'$ for small $a, \ a'$.


\section{Scheme Transformations for which $J(y)=(d/dy)\ln h(y)$}
\label{logderiv_section}

\subsection{General}
\label{logderiv_general_section}

In order for the general integral formalism that we have presented above to be
optimally useful, it is necessary that one should be able to do the integral
(\ref{fjint}) in closed form.  It is therefore helpful to consider a class of
Jacobian functions for which one is guaranteed to be able to calculate the
integral (\ref{fjint}).  Clearly, if $J(y)$ is the derivative of another
function, then one can always perform this integral. In this section we present
one such class of Jacobian functions.  These are functions that can be
expressed as logarithmic derivatives (LDs) of smooth functions denoted $h(y)$:
\beq
J(y) = \frac{d}{dy} \ln h(y) = \frac{h'(y)}{h(y)} \ , 
\label{jdlogh}
\eeq
where $h'(y) \equiv dh(y)/dy$. 
To obtain acceptable scheme transformation functions, we require that 
$h(y)$ is positive for physical (nonnegative) values of the argument $y$ and
that
\beq
h(0) = h'(0) \ . 
\label{hconstraint}
\eeq
The equality (\ref{hconstraint}) 
guarantees that the present construction satisfies the condition 
(\ref{j01}) that $J(0)=1$ and, as will be shown below, that it also satisfies
the condition (\ref{c0}) that $f(0)=1$. 
With $J(y)$ as specified in Eq. (\ref{jdlogh}), we can perform the 
integral (\ref{fjint}) immediately, obtaining the transformation function 
\beq
a = F(a')=\int_0^{a'} \, \frac{h'(y)}{h(y)} \, dy = 
\ln \Big [ \frac{h(a')}{h(0)} \Big ] \ . 
\label{Fjlogderiv}
\eeq
This shows why we required that $h(y)$ be positive for physical values of $y$,
since otherwise $h(0)$ and/or $h(a')$ might vanish, rendering the logarithm
singular. Since only the ratio $h(a')/h(0)$ enters in $F(a')$, it follows
that $F(a')$ is invariant under a rescaling of $h(y)$.  Consequently, we can,
without loss of generality, rescale $h(y)$ so that $h(0)=1$, and we shall do
this. Combining this with Eq. (\ref{hconstraint}), we have
\beq
h(0)=h'(0)=1  \ , 
\label{hconstraint2}
\eeq
and combining Eq. (\ref{hconstraint2}) with Eq. (\ref{Fjlogderiv}), we
obtain 
\beq
F(a') = \ln[h(a')] \ . 
\label{Fjlogderivn}
\eeq

To prove that this construction satisfies the condition $f(0)=1$, we 
use the definition (\ref{aap}) together with the analyticity of
$h(a')$ at $a'=0$.  We write out the Taylor series expansion for $h(a')$ at the
origin and use the property (\ref{hconstraint}) that we have imposed:
\beq
h(a') = 1 + a' + \frac{1}{2!} \, h''(0) \, (a')^2 + ... 
\label{htaylor} 
\eeq
where here and below, the dots $...$ denote higher powers of $a'$. 
Therefore, 
\beqs
f(0) & = & \lim_{a' \to 0} \, \frac{F(a')}{a'} \cr\cr
     & = & \lim_{a' \to 0} \, \frac{1}{a'} \ln \Big [ 1 + a' + 
\frac{1}{2} \, h''(0) \, (a')^2 + ... \Big ] \cr\cr
& = & 1 \ . 
\label{f01_logderiv}
\eeqs
Secondly, as noted above, this construction satisfies the condition $J(0)=1$. 
Since $a=F(a')$ by Eq. (\ref{aap}), Eq. (\ref{Fjlogderivn}) is equivalent to 
$e^a = h(a')$, so the inverse transformation is given formally as 
\beq
a' = h^{-1}(e^a) \ , 
\label{aprime__logderiv}
\eeq
where $h^{-1}$ denotes the inverse of the function $h$. We have found several
cases where this inverse can be calculated explicitly. We present some of these
next. 


\subsection{LD Function 1}
\label{ldogderiv1_section}

Here we present our first function $h$ to be used in 
Eq. (\ref{jdlogh}) and (\ref{Fjlogderiv}) to generate a new scheme
transformation. This is 
\beq
h(y) = 1 + \frac{1}{r}\ln(1+ry)  \ .  
\label{h_logderiv1}
\eeq
Hence, $h'(y) = 1/(1+ry)$.  By construction, this satisfies the condition 
that both $h(y)$ and $h'(y)$ are positive functions for physical (i.e.,
nonnegative) $y$ and the condition that $h(0)=h'(0)=1$. 
From (\ref{Fjlogderiv}), we have
\beq
a = F(a') = \ln \Big [ 1 + \frac{1}{r} \ln(1+ra') \Big ] \ . 
\label{F_logderiv1}
\eeq

We remark that for the families of scheme transformations studied so far in
\cite{scc}-\cite{sch3} that are dependent on an auxiliary parameter $r$, such
as $a=(1/r)\sinh(ra')$ and the transformations studied in \cite{schl} such as
$a=(1/r)\ln(1+ra')$ and $a=a'/(1-ra')$, setting $r=0$ yields the identity
transformation $a=F(a')=a'$.  However, this is not the case for the
transformation of Eq. (\ref{F_logderiv1}).  Instead, setting $r=0$ in
(\ref{F_logderiv1}) yields the scheme transformation \cite{rslr}
\beq
r=0 \quad \Longrightarrow \ \quad a = F(a') = \ln(1+a') \ . 
\label{r0aap}
\eeq
This property also holds for the transformations (\ref{F_logderiv2}) and
(\ref{F_logderiv3}) discussed below.  As is necessary, Eq. (\ref{r0aap}) 
obeys the requirement (\ref{c0}) that $f(0)=1$, i.e.,
that the transformation becomes an identity $a=a'$ in the free-field limit 
$a \to 0$.

The resultant series expansion for $f(a')=F(a')/a'$ has the form of
Eq. (\ref{fapseries}) with the $k_s$ coefficients displayed in Table
\ref{ks_table}.  In the special case $r=0$, the coefficients $k_s$ are given by
the Taylor series expansion of $(1/a')\ln(1+a')$ around $a'=0$, namely
\beq
r=0 \quad \Longrightarrow \ k_s = \frac{(-1)^s}{s+1} \ .  
\label{ks_r0}
\eeq
This is also true of the coefficients $k_s$ for the functions discussed in the
next two subsections, \ref{logderiv2_section} and 
\ref{logderiv3_section}.

The inverse transformation is
\beq
a' = \frac{1}{r} \Big [ \exp[r(e^a-1)]-1 \Big ] \ . 
\label{aprime_logderiv1}
\eeq

For the Jacobian, expressed in terms of $a'$ and $a$, we calculate 
\beqs
J & = & \frac{1}{(1+ra')\Big [ 1 + \frac{1}{r}\ln(1+ra') \Big ] } \cr\cr
  & = & \exp[-a-r(e^a-1)] \ . 
\label{j_logderiv1}
\eeqs
The parameter $r$ is restricted to the range 
\beq
r > -\frac{1}{a'}
\label{rrange_ld1}
\eeq
in order to avoid singularities in $h(y)$ and $F(a')$ and is further restricted
by the requirement that this scheme transformation must satisfy the conditions
C$_1$-C$_4$. These conditions can be satisfied for small positive $r$.  With
$r$ positive (indeed with $r > -1$), $k_{s_{min}} < 0$ (where $s_{min}=1$
here), so our general result (\ref{ks_aaprel}) implies that 
$a < a'$ for small $a, \ a'$. 


\subsection{LD Function 2}
\label{logderiv2_section}

As an input for the construction of our next new scheme transformation, we use 
\beq
h(y) = 1 + \frac{1}{r}\sinh(ry) \ .  
\label{h_logderiv2}
\eeq
Thus, $h'(y) = \cosh(ry)$.  Without loss of generality, the parameter $r$ can
be taken to be nonnegative, and we shall do this. 
Evidently, this function $h(y)$ satisfies the condition 
(\ref{hconstraint2}).  From the general result 
(\ref{Fjlogderiv}), we obtain
\beq
a = F(a') = \ln \Big [ 1 + \frac{1}{r} \sinh(ra') \Big ] \ . 
\label{F_logderiv2}
\eeq
The resultant series expansion for $f(a')=F(a')/a'$ has the form of
Eq. (\ref{fapseries}), and we list the first few coefficients $k_s$ in Table
\ref{ks_table}.  The invariance of the transformation $F(a')$ in
Eq. (\ref{F_logderiv2}) under the reversal in sign of the auxiliary parameter
$r$ is reflected in the property that the $k_s$ involve only even powers of
$r$.  Here $s_{min}=1$ and 
$k_{s_{min}} < 0$, so by our general result (\ref{ks_aaprel}), it follows that
$a < a'$ for small $a, \ a'$. 

The inverse transformation is
\beq
a' = \frac{1}{r} \ln \Big [ r(e^a-1)+ \sqrt{1+[r(e^a-1)]^2} \ \Big ] \ . 
\label{aprime_logderiv2}
\eeq
for the Jacobian we calculate 
\beq
J = \frac{\cosh(ra')}{1+ \frac{1}{r} \sinh(ra')} \ .
\label{j_logderiv2}
\eeq
This transformation satisfies all of the conditions C$_1$-C$_4$.  


\subsection{LD Function 3}
\label{logderiv3_section}

Here we discuss a third function $h$ for use in
Eq. (\ref{jdlogh}) and (\ref{Fjlogderiv}), namely 
\beq
h(y) = 1 + \frac{1}{r}\tanh(ry) \ . 
\label{h_logderiv3}
\eeq
Thus, $h'(y)=1/\cosh^2(ry)$.  This satisfies the condition
(\ref{hconstraint2}). As with the previous $h(y)$ function in
Eq. (\ref{h_logderiv2}), we can, without loss of generality, take the 
parameter $r$ to be nonnegative, and we shall do this. 
From the general result (\ref{Fjlogderiv}), we obtain
\beq
a = F(a') = \ln \Big [ 1 + \frac{1}{r}\tanh(ra') \Big ] \ . 
\label{F_logderiv3}
\eeq
The resultant series expansion for $f(a')=F(a')/a'$ has the form of
Eq. (\ref{fapseries}) with the first few $k_s$ coefficients listed in 
Table \ref{ks_table}. Since $s_{min}=1$ and $k_1 < 0$, we infer
that $a < a'$ for small $a, \ a'$. 

The inverse transformation is
\beq
a' = \frac{1}{2r} \, \ln \bigg [ \frac{1+r(e^a-1)}{1-r(e^a-1)} \bigg ] \ .
\label{aprime_logderiv3}
\eeq
The Jacobian, expressed as a function of $a'$ and of $a$, is 
\beqs
J(a')&=& \frac{1}{\cosh^2(ra') \, \Big [ 1 + \frac{1}{r}\tanh(ra') \Big ] } 
\cr\cr
& = & e^{-a}\Big [ 1+r(e^a-1)\Big ] \Big [ 1-r(e^a-1) \Big ] \ . \cr\cr
& & 
\label{j_logderiv3}
\eeqs
Although the transformation in Eq. (\ref{F_logderiv3}) and the Jacobian in 
Eq. (\ref{j_logderiv3}) are nonsingular for any
$r$, the inverse transformation (\ref{aprime_logderiv3}) 
does contain a singularity which restricts the
range of $r$.  Recalling that, without loss of generality, $r$ has been taken
to be nonnegative, this singularity occurs at $r=1/(e^a-1)$.  Hence, we
restrict $r$ to be substantially less than $1/(e^a-1)$ to avoid this 
singularity in the inverse transformation. 


\section{Scheme Transformations for which $J(y)=(d/dy)e^{\phi(y)}$}

\subsection{General} 

Here we present another class of $J(y)$ functions that can be used in
conjunction with our integral formalism to construct scheme transformations.
As was true of the functions in Section \ref{logderiv_section}, these function
have the form of total derivatives, which guarantees that one can do the
integral (\ref{fjint}).  We begin with an analytic function $\phi(y)$ that
satisfies the conditions
\beq
\phi(0) = 0, \quad \phi'(0)=1 \ .
\label{phiconditions}
\eeq
We then set $J(y)$ equal to the derivative of the exponential of this function:
\beq
J(y) = \frac{d}{dy} \, e^{\phi(y)} = \phi'(y) e^{\phi(y)} \ . 
\label{jphi}
\eeq
Substituting this into the integral (\ref{fjint}), we obtain
\beq
a = F(a') = e^{\phi(a')} - e^{\phi(0)} = e^{\phi(a')} - 1 \ . 
\label{Fphi}
\eeq
This yields $J(a') = dF(a')/da' = \phi'(a')e^{\phi(a')}$ so that, taking into
account the property (\ref{phiconditions}), it follows that $J(0)=1$.  
Furthermore, this construction guarantees that the condition $f(0)=1$ in 
Eq. (\ref{c0}) is satisfied.  To prove this, we use the defining relation 
$a=a'f(a')=F(a')$ in Eq. (\ref{aap}) to obtain 
\beq
f(a') = \frac{e^{\phi(a')}-1}{a'} \ .  
\label{fap_exp}
\eeq
Expanding the numerator in a Taylor series around $a'=0$, we get
\beqs
f(a') & = & \frac{1}{a'} \Big [ 
e^{\phi(0)} - 1 + \phi'(0)a' + O((a')^2) \Big ] \cr\cr
      & = & 1 + O(a') \ , 
\label{fapexpansion_exp}
\eeqs
from which it follows that $f(0)=1$.  The inverse is, formally, 
\beq
a' = \phi^{-1}[\ln(a+1)] \ , 
\label{apinverse_exp}
\eeq
where here $\phi^{-1}$ denotes the function that is the inverse of $\phi$.


\subsection{$\phi$ Function 1} 

In order to show how Eqs. (\ref{jphi}) and (\ref{Fphi}) can be used to
construct new scheme transformations, we first take
\beq
\phi(y) = \frac{1}{r}(e^{ry}-1) \ . 
\label{phi_exp1}
\eeq
For convenience, we may restrict $r$ to be nonnegative. 
This function satisfies the condition (\ref{phiconditions}). 
Substituting the resultant $J(y)=e^{ry}\exp[(1/r)(e^{ry}-1)]$ into 
Eq. (\ref{fjint}), we obtain the scheme transformation 
\beq
a = F(a') = \exp \Big [ \frac{1}{r}(e^{ra'}-1) \Big ] - 1 \ . 
\label{Fexp1}
\eeq
The resultant series expansion for $f(a')=F(a')/a'$ has the form of
Eq. (\ref{fapseries}) with the first few $k_s$ coefficients listed in Table
\ref{ks_table}. Note that in the limit as $r \to 0$, the scheme
transformation (\ref{Fexp1}) becomes 
\beq
r= 0 \quad \Longrightarrow \quad  a = F(a') = e^{a'}-1 \ .
\label{Fexp1_r0}
\eeq
Hence, in this limit the coefficients are given by 
\beq
r=0 \quad \Longrightarrow \quad k_s = \frac{1}{s!} \ .
\label{ks_exp1_r0}
\eeq
These results also hold for the transformation (\ref{Fexp2}) to be discussed
below. 

The inverse transformation is 
\beq
a' = \frac{1}{r} \ln \Big [ 1 + r\ln(a+1) \Big ] \ . 
\label{ap_exp1}
\eeq
Using this, we may express the Jacobian in terms of $a$:
\beqs
J & = & e^{ra'} \exp \Big [ \frac{1}{r}(e^{ra'}-1) \Big ] \cr\cr
  & = & (a+1)[1+r\ln(a+1)]  \ . 
\label{j_exp1}
\eeqs

This transformation satisfies conditions C$_1$-C$_4$.  


\subsection{$\phi$ Function 2}

As a second application of Eqs. (\ref{jphi}) and (\ref{Fphi}), we use
\beq
\phi(y) = \frac{1}{r} \sinh(ry) \ . 
\label{phi_exp2}
\eeq
Without loss of generality, we take the auxiliary parameter $r$ to
be nonnegative. This function satisfies the condition (\ref{phiconditions}). 
Substituting the resultant $J(y)= (d/dy) \, e^{(1/r)\sinh(ry)}$ into 
Eq. (\ref{fjint}), we obtain the scheme transformation 
\beq
a = F(a') = \exp\Big [ \frac{1}{r}\sinh(ra') \Big ] -1 \ . 
\label{Fexp2}
\eeq
The resultant series expansion for $f(a')=F(a')/a'$ has the form of
Eq. (\ref{fapseries}) with the $k_s$ coefficients listed in Table
\ref{ks_table}.  Because $k_{s_{min}} > 0$ (with $s_{min}=1$ here), 
our general result (\ref{ks_aaprel}) implies that $a > a'$ for small 
$a, \ a'$. 

The inverse transformation is 
\beq
a' = \frac{1}{r} \ln \Big [ r\ln(a+1) + \sqrt{1+ [r\ln(a+1)]^2} \ \Big ] \ . 
\label{ap_exp2}
\eeq
The Jacobian is 
\beq
J=\cosh(ra') \, \exp \Big [\frac{1}{r}\sinh(ra') \Big ] \ . 
\label{j_exp2}
\eeq
This transformation satisfies conditions C$_1$-C$_4$.  


\section{Comparative Analysis} 
\label{comparison_section}

In earlier work \cite{scc}-\cite{schl}, a number of scheme transformations have
been applied to ascertain the degree of scheme dependence of the value
$\alpha_{IR,n\ell}$ of the IR zero, calculated up to four-loop order, of the
beta function in an SU($N$) gauge theory with various fermion contents.
Comparisons have been made between results calculated in different schemes such
as $\overline{\rm MS}$, MOM, and ${\rm RI}'$,
\cite{tr,gracey2015,tr2016}. Scheme transformations have also been applied to
study the possibility of a UV zero in the beta function of a U(1) gauge theory
with $N_f$ (charged) fermions and in a globally invariant O($N$) $\lambda
|{\vec \phi}|^4$ theory up to the five-loop level \cite{lnf,lam}.

With the new scheme transformations generated by our integral formalism, we now
have a reasonably large set of such transformations to use to study scheme
dependence of the zero of a beta function away from the origin in
coupling-constant space.  In this section we include some remarks concerning
the analytic structure of these transformations that are relevant to this
application.  First, in the perturbative regime of small to moderate values of
$\alpha$ and hence also $\alpha'$ (which correspond to even smaller values of
$a=\alpha/(4\pi)$ and $a'=\alpha'/(4\pi)$), the effect of the scheme
transformation is largely determined by the values of the first few
coefficients $k_s$ in the Taylor series expansion of the transformation
function $f(a')$ in Eq. (\ref{fapseries}) or equivalently, $F(a')$ in Eq.
(\ref{Fapseries}) for the first few values of $s$.  Clearly, the same is true
of the inverse scheme transformation, as is evident from the Taylor series
(\ref{Fap_inverse_series}) for this inverse, together with the coefficients
$\rho_s$ determined via series reversion from the coefficients
$k_s$. Therefore, in this regime of moderately small couplings $\alpha$ and
$\alpha'$, one can get a reasonably good determination of the shift in the
value of $\alpha_{IR,n\ell}$ by examining the first few $k_s$ coefficients.  We
have listed these for comparative purposes in Table \ref{ks_table}.  The first
four lines of this table describe scheme transformations from
\cite{scc}-\cite{schl}, while the next seven lines describe new scheme
transformations presented and analyzed in the present paper.

Indeed, for small $a$ and hence also small $a'$, the question of whether a
given scheme transformation increases or decreases the coupling is determined,
using our general result (\ref{ks_aaprel}), by the sign of the lowest nonzero
coefficient, $k_{s_{min}}$, in the Taylor series expansion (\ref{fapseries}) or
equivalently, (\ref{Fapseries}), using our general result .  One can
conveniently read this from our Table \ref{ks_table} for the new scheme
transformations that we have presented in this paper. 

Hence, by combining the numerical analyses in \cite{scc}-\cite{schl} with the
analytic results for the first few $k_s$ coefficients in Table \ref{ks_table},
we can infer the effects of our new scheme transformations.  In particular, we
may again infer that in an asymptotically free non-Abelian gauge theory with a
two-loop IR zero at a value $\alpha_{IR,2\ell}$ that is not too large, and for
moderate values of the auxiliary parameter $r$, the scheme dependence inherent
in the calculation of $\alpha_{IR,n\ell}$ at $n=3$ and $n=4$ loops is
moderately small.


\section{Conclusions}
\label{conclusions} 

In this paper we have presented an integral formalism for constructing scheme
transformations.  We have used this formalism to generate several new scheme
transformations that are acceptable for the analysis of a zero of the beta
function away from the origin in coupling-constant space.  By performing
Taylor-series expansions of these scheme transformations, we have formulated an
analytic approach to their effect on the coupling. These results bolster
previous numerical studies to show that in an asymptotically free gauge theory
with an two-loop value of the IR zero of the beta function,
$\alpha_{IR,2\ell}$, that is not too large, the scheme
transformations presented here (with moderate values of the auxiliary parameter
$r$) produce only relatively mild shifts in higher-loop values
$\alpha_{IR,n\ell}$.  


\begin{acknowledgments}

This research was partly supported by the NSF Grant No. NSF-PHY-13-16617. 

\end{acknowledgments}


\begin{appendix}

\section{A Result from Calculus} 
\label{diffint_appendix}

We use the following result from multivariable calculus.  Consider the integral
\beq
F(x) = \int_{y_1(x)}^{y_2(x)} \Phi(x,y) \, dy \ . 
\label{fxint}
\eeq
Then,
\beqs
\frac{dF(x)}{dx}&=& \frac{dy_2(x)}{dx} \, \Phi(x,y_2(x)) - 
                    \frac{dy_1(x)}{dx} \, \Phi(x,y_1(x)) \cr\cr
& + & \int_{y_1(x)}^{y_2(x)} \frac{\partial \Phi(x,y)}{\partial x} \, dy \ . 
\label{diffint}
\eeqs
In particular, if $\Phi(x,y)$ does not depend on $x$, which we indicate by 
setting $\Phi(x,y) \equiv J(y)$ (which may depend on auxiliary parameters such
as $r$), and if $y_2(x)=x$ and $y_1(x)={\rm const.}$, then 
Eq. (\ref{diffint}) reduces to $dF/dx=J(x)$. 


\end{appendix}



\newpage
\begin{widetext}
\begin{table}
\caption{\footnotesize{Values of the coefficients $k_s$ in
Eq. (\ref{fapseries}) for scheme transformations discussed in the text. 
The transformation $F(a')$ is defined by Eq. (\ref{aap}): $a=a'f(a')=F(a')$.
The equation numbers indicate where the given $F(a')$ is presented in the
text.}}
\begin{center}
\begin{tabular}{|c|c|c|c|c|c|} \hline\hline
$F(a')$ & Eq. & $k_1$ & $k_2$ & $k_3$ & $k_4$ \\
\hline
$(1/r)\sinh(ra')$  & (\ref{st_sinh}) & 0  & $r^2/6$  & 0  & $r^4/120$  \\
\hline
$(1/r)\tanh(ra')$  & (\ref{st_tanh}) &  0  & $-r^2/3$ & 0  & $2r^4/15$   \\
\hline
$(1/r)\ln(1+ra')$  & (\ref{slr}) & 
$-r/2$ & $r^2/3$ & $-r^3/4$ & $r^4/5$ \\
\hline
$a'/(1-ra')$ & (\ref{st_rational}) & $r$ & $r^2$ & $r^3$ & $r^4$ \\
\hline
$\frac{(1+ra')^{p+1}-1}{r(p+1)}$ & (\ref{F_alg}) & 
$\frac{r}{2} \, p$ & 
$\frac{r^2}{3}\, {p \choose 2}$ & $\frac{r^3}{4}\, {p \choose 3}$ & 
$\frac{r^4}{5}\, {p \choose 4}$ \\
\hline
$a'+(1/r)\ln[\cosh(ra')]$ & (\ref{F_tran}) & 
$r/2$ & 0 & $-r^3/12$ & 0 \\
\hline
$\ln[1+(1/r)\ln(1+ra')]$ & (\ref{F_logderiv1}) & 
$-(r+1)/2$ & $(2r^2+3r+2)/6$ & $-(6r^3+11r^2+12r+6)/24$
& $(12r^4+25r^3+35r^2+30r+12)/60$ \\
\hline
$\ln[1+(1/r)\sinh(ra')]$ & (\ref{F_logderiv2}) & 
$-1/2$ & $(r^2+2)/6$ & $-(2r^2+3)/12$ & 
$(r^4+20r^2+24)/120$ \\ 
\hline
$\ln[1+(1/r)\tanh(ra')]$ & (\ref{F_logderiv3}) & 
$-1/2$ & $(1-r^2)/3$ & $(4r^2-3)/12$ &
$(r^2-1)(2r^2-3)/15$ \\
\hline
$\exp[(1/r)(e^{ra'}-1)]-1$ & (\ref{Fexp1}) & 
$(r+1)/2$ & $(r^2+3r+1)/6$ & 
$(r^3+7r^2+6r+1)/24$ & $(r^4+15r^3+25r^2+10r+1)/120$ \\ 
\hline
$\exp[(1/r)\sinh(ra')]-1$ & (\ref{Fexp2}) & 
$1/2$ & $(r^2+1)/6$ & $(4r^2+1)/24$ & 
$(r^4+10r^2+1)/120$ \\
\hline\hline
\end{tabular}
\end{center}
\label{ks_table}
\end{table}
\end{widetext}



\begin{thebibliography}{99}
\bibitem{rslr}
%
Note that this is a special case of the scheme transformation (\ref{slr})
that we studied in \cite{schl} with the (different) auxiliary parameter $r$ in
Eq. (\ref{slr}) set equal to 1.
\end{thebibliography}
\end{document}